\pdfoutput=1             

\documentclass[preprint,12pt,authoryear]{elsarticle}

\makeatletter
\def\ps@pprintTitle{%
   \let\@oddhead\@empty
   \let\@evenhead\@empty
   \let\@oddfoot\@empty
   \let\@evenfoot\@oddfoot}
\makeatother

\usepackage{setspace} 
\usepackage{lineno} 
\usepackage[nohyperlinks]{acronym}
\usepackage[hidelinks]{hyperref}

\usepackage{wrapfig} 
\usepackage{lineno} 

\usepackage{tikz}

\usetikzlibrary{shapes.geometric, arrows} 
\usepackage{pdfpages}
\usepackage[T1]{fontenc}
\usepackage[utf8]{inputenc}
\usepackage{lmodern}
\usepackage{amsmath,amssymb,mathtools}
\usepackage[capitalise, noabbrev]{cleveref} 
\usepackage{graphicx}
\usepackage{booktabs}
\usepackage{siunitx}
\usepackage{doi}
\usepackage{listings}
\usepackage{caption}

\usepackage{subcaption} 
\usepackage{natbib}
\usepackage{doi} 

\usepackage{enumitem}

\usepackage[section]{placeins} 

\begin{document}

\doublespacing
\acresetall
\title{Routing-Method Effects on Distance, Time, Fuel, and Emissions in Europe–Asia Trade: A Comparison of the Suez, Cape, and Northern Sea Route Corridors}

\author[1,2]{Abdella Mohamed\corref{cor1}}
\ead{mohamed.abdella@tum.de}

\author[2]{Christian Hendricks}
\ead{christian.hendricks@everllence.com}

\author[1]{Xiangyu Hu\corref{cor1}}
\ead{xiangyu.hu@tum.de}

\cortext[cor1]{Corresponding author}

\affiliation[1]{%
  organization={Technical University of Munich},
  addressline={Boltzmannstraße 15},
  postcode={85748},
  city={Garching},
  country={Germany}
}

\affiliation[2]{%
  organization={Everllence (formerly: MAN Energy Solutions)},
  addressline={Stadtbachstraße 1},
  postcode={86153},
  city={Augsburg},
  country={Germany}
}


\begin{frontmatter}

\begin{abstract}

Growing interest in decarbonization and Arctic accessibility has renewed interest in Europe–Asia shipping corridors. The Northern Sea Route (NSR) is often portrayed as a 30--40\% shortcut relative to Suez, with savings propagated to time, fuel, and CO$_2$. The effect of enforcing sea-only feasibility on
these baselines, and its downstream impact on time, fuel, and CO2, remains
under-examined.

We compare great-circle baselines with sea-only routes computed via A-star search (A*) on a 0.5$^\circ$ grid between Northern Europe and Northeast Asia across the Suez, Cape of Good Hope, and NSR corridors under three waypoint philosophies. Distances are mapped to voyage time using corridor-typical speeds and to fuel/CO$_2$ using main- and auxiliary-engine accounting.

Sea-only routing preserves the ranking NSR $<$ Suez $<$ Cape but compresses NSR’s advantage once realistic speeds are applied. NSR remains shortest (about 8--10k~nm versus 11--12k~nm for Suez), yet typical durations differ modestly and fuel/CO$_2$ savings over Suez are small and variant-dependent. Equal-speed tests restore geometric ordering, and endpoint sensitivity shows larger NSR gains for more northern East Asian ports.

The framework provides a reproducible, corridor-agnostic benchmark for later integration of sea ice, weather, regulatory overlays, and AIS data in dynamic Arctic voyage planning.

\end{abstract}


\end{frontmatter}

\section*{Nomenclature}
\renewcommand{\baselinestretch}{0.75}\normalsize
\renewcommand{\aclabelfont}[1]{\textsc{\acsfont{#1}}}
\begin{acronym}[longest]

    \acro{a-star}[A*]{A-star Pathfinding Algorithm}
    \acro{ae}[AE]{Auxiliary Engine}
    \acro{ais}[AIS]{Automatic Identification System}

    \acro{co2}[CO\textsubscript{2}]{Carbon Dioxide}

    \acro{dwt}[DWT]{Deadweight Tonnage}

    \acro{gc}[GC]{Great Circle}
    \acro{ghg}[GHG]{Greenhouse Gas}

    \acro{imo}[IMO]{International Maritime Organization}

    \acro{me}[ME]{Main Engine}

    \acro{nsr}[NSR]{Northern Sea Route}

    \acro{pm}[PM]{Particulate Matter}

    \acro{suez}[Suez]{Suez Canal}

    \acro{tss}[TSS]{Traffic Separation Schemes}

\end{acronym}
\renewcommand{\baselinestretch}{1}\normalsize

\section{Introduction}\label{sec:introduction}

Growing attention to decarbonization and Arctic accessibility has renewed interest in alternative Europe--Asia shipping corridors. The Northern Sea Route (NSR) is frequently presented as a 30--40\% shorter alternative to Suez based on great-circle (GC) geometry \citep{liu2010_nsr,furuichi2015_platform,schoyen2011_nsrecon}. Recent corridor assessments continue to report NSR advantages primarily from geometric or stylized baselines before adding cost or feasibility layers \citep{li2023_transarctic_jmse,zeng2020_competitiveness}. Some studies propagate these geometric savings to time, fuel, or CO$_2$ without first demonstrating sea-only navigability \citep{iaph2013_portplanning,wan2018_sustainability}, whereas others explicitly model Arctic feasibility and operating speeds, often jointly with regulatory or emissions-control layers \citep{SmithStephenson2013,melia2016_future,kavirathna2023_nsr_feasibility}. Because corridor baselines are increasingly used to parameterize network, climate, and policy models, small routing-representation biases can scale into large system-level conclusions \citep{poo2024_arctic_network}. What remains under-examined is how enforcing sea-only feasibility reshapes corridor baselines and their downstream translation to time, fuel, and CO$_2$ under corridor-specific speeds.

We compare geometric GC baselines (direct GC and GC via corridor macro-waypoints) with physically feasible sea-only routes computed on a coastline-masked 0.5$^\circ$ A* graph. The framework is applied to Rotterdam--Yokohama, a representative Europe--Asia pair spanning the Suez, Cape of Good Hope, and NSR corridors. Each corridor is represented by three waypoint variants to capture distinct routing philosophies: (i) \emph{service-guided} (hub/chokepoint oriented), (ii) \emph{bluewater-oriented} (open-ocean favouring), and (iii) \emph{channel/coast-guided} (lane/coast following). This design magnifies routing-method effects over an intercontinental distance while remaining comparable to prior corridor studies \citep{li2023_transarctic_jmse,kavirathna2023_nsr_feasibility,meza2023_arctic_routes}. Distances are mapped to voyage time using corridor-typical service speeds and to fuel/CO$_2$ using explicit main- and auxiliary-engine accounting. To isolate routing effects, we exclude sea ice, weather, emission control areas (ECAs), bathymetry and draft limits, and AIS-based tracking; these layers are reserved for subsequent constrained analyses so that environmental and regulatory factors do not confound the methodological comparison.

Our analysis yields three headline insights. First, enforcing sea-only feasibility preserves the distance ranking (NSR $<$ Suez $<$ Cape): NSR remains shortest once routes are rendered navigable. Second, distance does not translate linearly to time or fuel---corridor-specific speeds and practices such as slow steaming can compress, or even eliminate, NSR's apparent operational advantage over Suez \citep{notteboom2009_fuelcosts,maloni2013_slowsteaming}. Third, although longest, the Cape corridor remains a robust year-round fall-back independent of canals and Arctic seasonality, consistent with the recent diversion of container services around Africa during the Red Sea crisis \citep{Notteboom2024MEL}.

We focus on Europe--Asia trade because it is a major deep-sea liner market and a key arena for debates on the competitiveness of alternative long-haul corridors. Rotterdam--Yokohama spans the Suez, Cape of Good Hope, and NSR corridors and is widely used as a representative origin--destination pair in the NSR--Suez literature. Recent route-comparison studies likewise use Rotterdam--Asia pairs to quantify Arctic versus Suez corridor differences \citep{meza2023_arctic_routes}. 

Macro-waypoints were placed at natural gateway regions (canal entrances, straits, capes). Each successive pair was verified to admit a valid sea-only path in \texttt{searoute}; cases with land intersections or graph failures were iteratively adjusted and then checked visually on interactive charts.

From the reviewed literature, three methodological gaps emerge:

\begin{enumerate}[label=(\roman*),leftmargin=2em]
  \item \textbf{Method isolation is rare.} Comparative studies often remain geometric (GC) or embed routing within complex environmental or economic models, obscuring how routing representation itself biases distance or derived efficiency metrics \citep{liu2010_nsr,schoyen2011_nsrecon,SmithStephenson2013,melia2016_future}.
  \item \textbf{Propagation into emissions is underexplored.} Few works trace routing assumptions through to fuel and CO$_2$ accounting with explicit treatments of main- and auxiliary-engine loads \citep{schroeder2017_emission,chen2021_passenger,karamperidis2022_review,johansson2022_uncertainty,nguyen2023_sensitivity}.
  \item \textbf{Reproducibility is uneven.} Many studies provide limited access to code, parameters, or scenario definitions, hindering cross-study benchmarking and robustness checks.
\end{enumerate}

The primary objective is to quantify how routing representation (GC versus sea-only A* on a 0.5$^\circ$ water mask) alters corridor distances and their translation to voyage time, fuel, and CO$_2$ for Rotterdam--Yokohama across Suez, Cape of Good Hope, and NSR. Secondary objectives are to contrast corridor-typical versus equal-speed baselines, test endpoint sensitivity (for example, Busan versus Yokohama). 

\noindent\textbf{Contributions.} This study provides:
\begin{enumerate}[label=(\alph*),leftmargin=2em]
  \item a corridor-agnostic, reproducible sea-only benchmark for Europe--Asia routing across Suez, Cape of Good Hope, and NSR;
  \item quantified GC$\rightarrow$sea-only deltas under three waypoint philosophies;
  \item transparent propagation to time and fuel/CO$_2$ using explicit main- and auxiliary-engine accounting; and
  \item robustness and endpoint sensitivity analyses identifying when NSR's geometric advantage does---and does not---translate into operational savings.
\end{enumerate}

\section{Data and Methods}\label{sec:data_methods}

\subsection{Study scope and corridor design}\label{sec:scope}

The analysis targets deep-sea services between Northern Europe and Northeast Asia, 
where the choice of long-haul corridor has the largest implications for distance, 
time, fuel, and emissions. We select Rotterdam (51.95$^{\circ}$N, 4.14$^{\circ}$E) 
and Yokohama (35.45$^{\circ}$N, 139.65$^{\circ}$E) as the baseline 
origin--destination pair because it:
\begin{itemize}
  \item lies on a major Europe--Asia trade lane,
  \item is widely used in comparative corridor studies, and
  \item can be served via Suez, Cape of Good Hope, or the Northern Sea Route (NSR).
\end{itemize}
This ensures both operational relevance and comparability to earlier work, while 
magnifying routing-method effects over intercontinental distance.

We explicitly include three strategic maritime corridors:
\begin{enumerate}[label=(\roman*)]
    \item \textbf{Suez Canal Corridor (SUEZ)}: the established Europe--Asia trunk 
    route via the Mediterranean, Suez Canal, Red Sea, and Indian Ocean;
    \item \textbf{Cape of Good Hope Corridor (CAPE)}: a canal-independent fallback 
    via the Atlantic and Southern Indian Oceans, used during Suez and Red Sea 
    disruptions;
    \item \textbf{Northern Sea Route (NSR)}: the high-latitude Arctic corridor 
    along the Russian coast between the Barents Sea and Bering Strait.
\end{enumerate}

\begin{figure}[t]
  \centering
  \includegraphics[width=\linewidth]{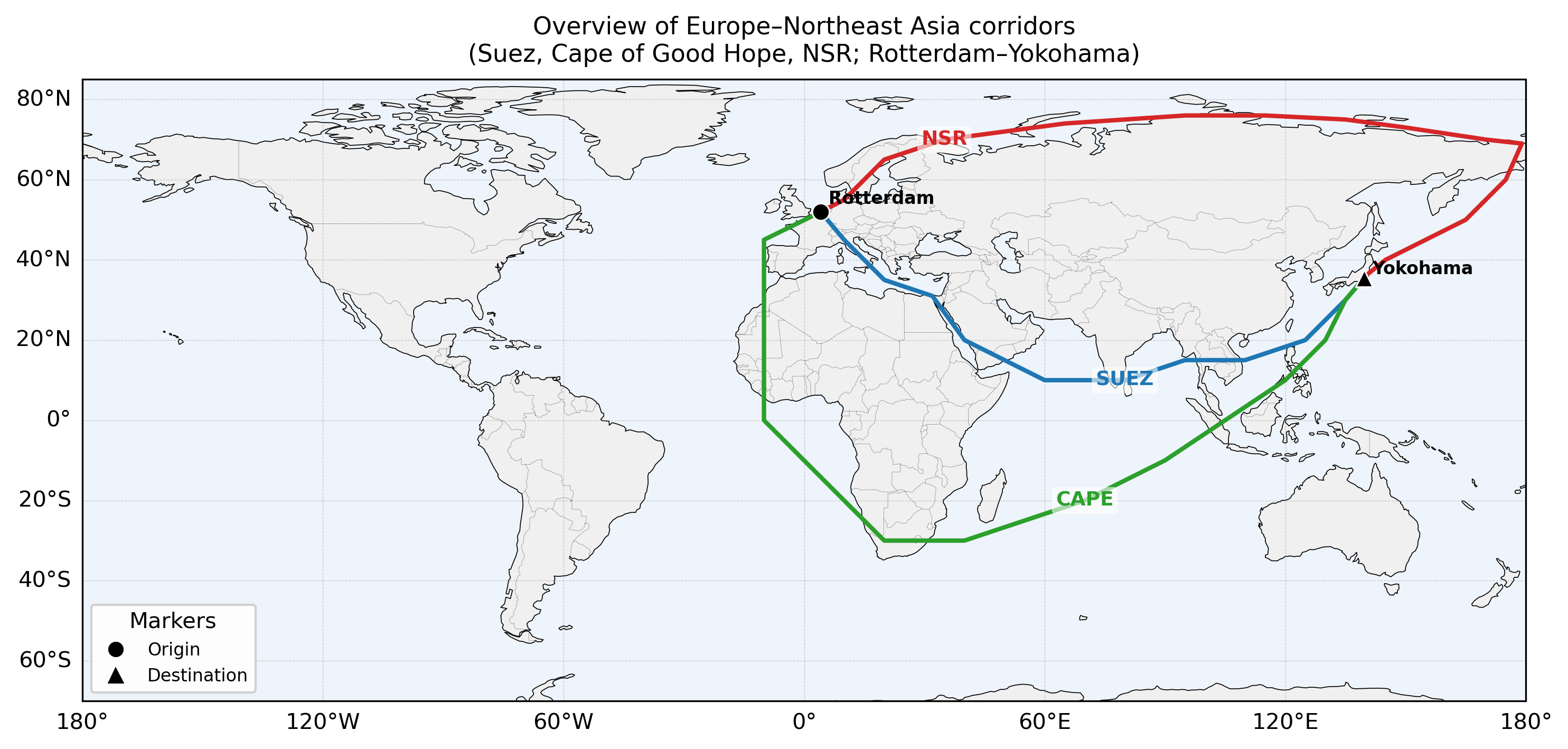}
  \caption{Overview of the three Europe–Northeast Asia corridors considered in this study}
  \label{fig:corridor_overview}
\end{figure}

A global overview of these three corridors for the Rotterdam–Yokohama origin–destination pair is shown in Figure~\ref{fig:corridor_overview}.

For context, the direct great-circle distance between Rotterdam and Yokohama is about 5.1 k nm, i.e. substantially shorter than any of the corridorised routes considered here.

\FloatBarrier

We do not include the Panama Canal or the Northwest Passage. For the selected 
Northern Europe--Northeast Asia trade, Panama is not a competitive option, and 
the Northwest Passage remains operationally rare and heavily constrained. 
Including them would add complexity without materially changing the main corridor 
competition for this trade.

For each corridor we define three operational \emph{waypoint philosophies}:
\begin{enumerate}[label=(\alph*)]
  \item a \emph{service-guided} variant that loosely follows typical liner 
  schedules and gateway regions (major hubs and chokepoints);
  \item a \emph{bluewater-oriented} variant favouring longer open-ocean legs 
  where feasible; and
  \item a \emph{channel/coast-guided} variant that adheres more closely to 
  constrained straits and coastal guidance.
\end{enumerate}
These nine corridor--variant combinations bracket plausible operator routing 
styles without requiring detailed commercial schedules. Waypoints were defined manually based on contemporary sailing directions 
and publicly available track patterns, then verified via visual inspection 
on interactive charts to ensure they represent realistic chokepoints and 
gateway regions rather than arbitrary mid-ocean points. They allow us to assess whether the 
effects of routing representation (geometric vs.\ sea-only) persist across 
reasonable corridor geometries. 
The macro waypoints and variant philosophies are summarized in Table~\ref{tab:waypoint_manifest}. 

\begin{table}[t]
  \centering
  \caption{Macro waypoint manifest and routing-variant philosophies for each corridor. 
  Variants correspond to (a) service-guided, (b) bluewater-oriented, and 
  (c) channel/coast-guided waypoint chains used to define corridor-specific GC and sea-only routes.}
  \label{tab:waypoint_manifest}
  \resizebox{\linewidth}{!}{%
    \begin{tabular}{lllrl}
\toprule
           scenario & corridor & philosophy &  n\_waypoints &                                    waypoints\_chain \\
\midrule
  CAPE\_VAR1\_service &     CAPE &       VAR1 &           10 & (52.0,4.1) → (36.0,-5.6) → (14.7,-17.5) → (-34.... \\
CAPE\_VAR2\_bluewater &     CAPE &       VAR2 &            9 & (52.0,4.1) → (30.0,-20.0) → (-10.0,-25.0) → (-3... \\
    CAPE\_VAR3\_coast &     CAPE &       VAR3 &           12 & (52.0,4.1) → (36.0,-5.6) → (5.0,-5.0) → (-20.0,... \\
   NSR\_VAR1\_service &      NSR &       VAR1 &            9 & (52.0,4.1) → (71.0,25.0) → (70.5,58.0) → (76.0,... \\
 NSR\_VAR2\_bluewater &      NSR &       VAR2 &            9 & (52.0,4.1) → (73.0,35.0) → (70.5,58.0) → (76.5,... \\
     NSR\_VAR3\_coast &      NSR &       VAR3 &            9 & (52.0,4.1) → (70.0,33.0) → (70.5,58.0) → (75.0,... \\
  SUEZ\_VAR1\_service &     SUEZ &       VAR1 &           10 & (52.0,4.1) → (50.5,1.0) → (36.0,-5.6) → (30.0,3... \\
SUEZ\_VAR2\_bluewater &     SUEZ &       VAR2 &           10 & (52.0,4.1) → (48.0,-10.0) → (36.0,-5.6) → (30.0... \\
    SUEZ\_VAR3\_coast &     SUEZ &       VAR3 &           13 & (52.0,4.1) → (50.5,1.0) → (36.0,-5.6) → (33.5,2... \\
\bottomrule
\end{tabular}

  }
\end{table}

\subsection{Routing representations: geometric GC vs.\ sea-only A*}\label{sec:routing}

We compare two routing representations for each corridor and waypoint variant.

\subsubsection{Great-circle baselines}\label{sec:gc}

Great-circle (GC) paths represent the geometric minimum distance between two 
points on a sphere. They are widely used in maritime and transport studies as 
distance benchmarks and as inputs to cost and emission models. However, raw GC 
arcs do not incorporate land--sea geometry, do not respect coastlines or 
chokepoints, and in principle may cross continental landmasses. In practice, 
GC-based corridor distances often reflect either (i) a single GC between origin 
and destination or (ii) a sum of GCs between a small set of corridor waypoints. 
In both cases, GC distances act as \emph{optimistic lower bounds} that idealise 
the Earth as a smooth sphere and abstract away navigational feasibility.

In this study, we construct for each corridor and waypoint variant a 
\emph{GC chain} connecting Rotterdam, the corridor-specific macro-waypoints, and 
Yokohama in sequence. These chains capture the geometric potential of each 
corridor and are representative of the kind of distances often invoked when the 
NSR is described as ``30--40\% shorter than Suez''.

\subsubsection{Sea-only routing with A*}\label{sec:astar}

To obtain physically feasible paths, we use the open-source \texttt{searoute} 
package \citep{searoute2022}, which implements a Dijkstra/A* shortest-path search on a global 
oceanic graph. Internally, \texttt{searoute} uses a regular latitude--longitude 
grid (approximately 0.5$^{\circ}$ resolution) whose nodes are constrained to 
water cells; coastlines and land are masked out when constructing the graph 
(as implemented in \texttt{searoute}'s default coastline mask). 
Edges connect neighbouring water nodes with weights proportional to 
great-circle segment length.

For each consecutive waypoint pair in a corridor's macro-waypoint chain, 
\texttt{searoute} is used to compute a sea-only shortest path. The resulting 
segments are concatenated to form a continuous sea-only polyline from Rotterdam 
to Yokohama. Every step of the search is constrained to water nodes; land and 
narrow straits act as hard constraints. This sea-only representation therefore captures corridor geometry, including 
high-latitude bends and mandatory chokepoints along each route (e.g.\ Bering 
Strait on the NSR, the Suez Canal, and the Cape of Good Hope), and provides an 
operationally meaningful approximation of feasible vessel trajectories. We verified algorithmic parity by confirming that A* and Dijkstra return identical shortest-path distances on the same sea graph.

The effective grid resolution (\mbox{$\approx 0.5^{\circ}$}) is chosen as a 
compromise between computational cost and geometric fidelity. At 
intercontinental scales, corridor-level differences in distance and time are 
insensitive to finer resolutions, whereas substantially coarser grids risk 
misrepresenting narrow passages or coastal stand-off. At 60$^{\circ}$N, a 
0.5$^{\circ}$ cell corresponds to roughly 15--30~nm spacing, adequate for corridor-scale benchmarking. 
Because our aim is to 
isolate \emph{routing-method effects} rather than resolve harbour approaches or 
traffic separation schemes (TSS), this resolution is sufficient.

Antimeridian crossings (near $\pm 180^{\circ}$ longitude) are handled through a 
combination of \texttt{searoute}'s internal graph representation and explicit 
dateline guards in post-processing, ensuring that polylines are continuous and 
free of wrap-around artefacts. Route geometries were validated by segment-length 
diagnostics to confirm no wrap-around jumps. Route polylines are subsequently interpolated 
to a uniform spacing (0.25$^{\circ}$) for consistent distance estimation.

\subsubsection{Rationale for comparing GC and A*}\label{sec:why_GC_Astar}

Comparing GC and A* isolates the effect of \emph{routing representation} itself: 
to what extent widely reported corridor advantages (e.g.\ ``NSR is 30--40\% 
shorter than Suez'') derive from geometric idealisations rather than navigable 
paths. GC chains bound the problem from below as idealised geometric minima. 
Sea-only A* routes represent the shortest navigable paths under static constraints, providing a conservative baseline prior to adding ice or weather costs.

Alternative graph-search or optimisation methods (e.g.\ Dijkstra on the same 
graph, dynamic programming, evolutionary heuristics, reinforcement learning) 
would return the same shortest path as A* under identical static constraints on 
a single-objective distance cost. Their added value lies in multi-objective or 
metocean-aware optimisation, which we deliberately reserve for subsequent work 
on dynamic voyage planning and environmental routing. The goal here is to 
cleanly separate the effect of moving from geometric GC representations to 
sea-only, coastline-respecting paths and to quantify how this transition 
propagates into distance, time, fuel, and CO$_2$ across the three corridors.

\subsection{From distance to time, fuel, and CO$_2$}\label{sec:energy_model}

For each scenario (corridor $\times$ waypoint philosophy $\times$ routing 
method), total route distance $D$ [nm] is computed as the sum of 
great-circle segment lengths along the GC chain or A* sea-only polyline.

Indicative voyage time is then obtained using corridor-specific average service 
speeds $U_{\text{corridor}}$ representative of current practice:
\begin{equation}
  T = \frac{D}{U_{\text{corridor}}},
\end{equation}
where $T$ is the sailing time. We do not optimise speeds; instead, we fix 
representative values to highlight how routing-method differences interact with 
realistic speed policies:
\begin{itemize}
    \item NSR: 12.5 kn (reflecting ice-class and high-latitude operational limits) \citep{shu2024_arctic_speed,li2024_nsr_speed_uncertainty};
    \item SUEZ: 14.5 kn (typical mainline container service under slow-steaming practice) \citep{bimco2024_container_outlook};
    \item CAPE: 14.0 kn (open-ocean service; conservative relative to Cape-rerouting averages) \citep{bimco2024_container_outlook}.
\end{itemize}
These values reflect corridor-typical slow-steaming practice and Arctic speed limits and are used to isolate routing-method effects rather than optimise operations \citep{bimco2024_container_outlook,shu2024_arctic_speed}.

Fuel consumption and CO$_2$ emissions are derived from voyage time using 
explicit main- and auxiliary-engine accounting. For the main engine, we assume a cubic speed--power relation and scale power from a reference operating point $(P_0,U_0)$:
\begin{equation}
  P_{\text{ME}} = P_0 \left(\frac{U}{U_0}\right)^3,
  \qquad 
  \dot{m}_{\text{ME}} = P_{\text{ME}} \cdot \mathrm{SFOC}_{\text{ME}},
\end{equation}
where $(P_0,U_0)$ is a reference condition (Table~\ref{tab:params}). 
SFOC values are converted from g\,kWh$^{-1}$ to t\,h$^{-1}$ using standard unit factors. 
Total main-engine fuel is $m_{\text{ME}} = \dot{m}_{\text{ME}} \, T$. 
Auxiliary-engine fuel is modelled as a constant power load over time,
$m_{\text{AUX}} = \dot{m}_{\text{AUX}} \, T$, capturing hotel and support loads 
during the voyage. The combined fuel consumption is then
\begin{equation}
  m_{\text{FUEL}} = m_{\text{ME}} + m_{\text{AUX}}.
\end{equation}
CO$_2$ emissions are computed using a standard emission factor 
$EF_{\text{CO2}}$ [t CO$_2$/t fuel]:
\begin{equation}
  m_{\text{CO2}} = m_{\text{FUEL}} \cdot EF_{\text{CO2}}.
\end{equation}

The cubic speed--power relation, constant SFOC, and CO$_2$ emission factor follow common practice in shipping 
emission inventories and IMO GHG studies \citep{emepeea_2019,imo_4th_ghg_2020,smith2014_third_ghg}. 
This level of detail is sufficient for isolating routing-method effects on 
voyage time, fuel, and CO$_2$; higher-fidelity resistance and engine 
models are reserved for subsequent work.

We intentionally retain a simple, transparent fuel and emission model so that 
the influence of routing representation on $T$, $m_{\text{FUEL}}$, and 
$m_{\text{CO2}}$ is fully traceable. Ship-specific resistance, 
weather-dependent added resistance, and speed-control strategies are left for 
follow-up work.

Representative parameter values (for a Panamax container vessel of 
approximately 50\,000--70\,000 DWT on conventional HFO) are summarised in 
Table~\ref{tab:params} and are consistent with standard 
practice in shipping emission inventories and IMO guideline assumptions 
\citep[e.g.][]{imo_4th_ghg_2020,emepeea_2019,smith2014_third_ghg}. 

A Panamax-class container vessel is used as a neutral baseline to convert voyage time into indicative fuel and CO$_2$. Absolute totals depend on ship size; however, under first-order scaling with common speed--power and SFOC assumptions, corridor rankings and relative routing-method deltas are expected to remain stable. Vessel-size sensitivity is examined in follow-on work. The parameter values are chosen to represent a generic Panamax vessel rather than a specific ship, so results reflect typical corridor-level behaviour instead of vessel-specific performance. The exact numerical values can be updated without changing the structure of the workflow.

\begin{table}[t]
  \centering
\caption{Assumed vessel, engine, and fuel parameters for the indicative 
time--fuel--CO$_2$ calculations (Panamax / small post-Panamax container 
vessel). Values can be adapted to other ship types without altering the methodology.}
  \label{tab:params}
  \begin{tabular}{lll}
    \toprule
    Symbol & Description & Value (example) \\
    \midrule
    $P_0$       & Reference main-engine power        & 35\,000 kW \\
    $U_0$       & Reference service speed            & 14.5 kn \\
    $\mathrm{SFOC}_\mathrm{ME}$ & Main engine SFOC        & 170 g\,kWh$^{-1}$ \\
    $P_\mathrm{AE}$      & Auxiliary power                & 2\,000 kW \\
    $\mathrm{SFOC}_\mathrm{AE}$ & Auxiliary SFOC          & 185 g\,kWh$^{-1}$ \\
    $EF_{\mathrm{CO_2}}$ & CO$_2$ factor (HFO)           & 3.114 t\,CO$_2$/t\,fuel \\
    $U_\mathrm{NSR}$     & Representative NSR speed       & 12.5 kn \\
    $U_\mathrm{SUEZ}$    & Representative Suez speed      & 14.5 kn \\
    $U_\mathrm{CAPE}$    & Representative Cape speed      & 14.0 kn \\
    \bottomrule
  \end{tabular}
\end{table}

All computations were performed in Python~3.11 using \texttt{pandas}, 
\texttt{geopy}, \texttt{shapely}, and \texttt{searoute}. Derived data tables were 
exported as CSV and \LaTeX{} tables, and visualised through static charts and 
interactive \texttt{folium} maps. 

\subsection{Scenario matrix and sensitivity tests}\label{sec:scenarios}

The primary scenario matrix consists of:
\begin{itemize}
  \item three corridors (SUEZ, CAPE, NSR);
  \item three waypoint philosophies per corridor (service-guided, 
  bluewater-oriented, channel/coast-guided);
  \item two routing representations (GC chain and A* sea-only); and
  \item corridor-specific average service speeds as in Table~\ref{tab:params}.
\end{itemize}
For each element in this matrix, we compute distance, indicative time, fuel, and 
CO$_2$ as described above.

To assess robustness, we perform several complementary sensitivity analyses:
\begin{enumerate}[label=(\alph*)]
  \item \textbf{Endpoint sensitivity:} repeating the analysis with an East Asia 
  endpoint in Busan instead of Yokohama to assess how NSR's relative advantage 
  changes for more northerly/westerly destinations.
  \item \textbf{Equal-speed comparisons:} repeating the time and fuel 
  calculations with a common service speed (14 kn) across all corridors to 
  isolate geometric and routing effects from speed policy.
\end{enumerate}

We deliberately exclude dynamic environmental and regulatory constraints (sea 
ice, winds, waves, currents, emission control areas, canal queues, ice 
pilotage/escort requirements) to maintain focus on routing-method effects. 
These factors will be layered onto the A* sea-only baseline in subsequent work 
on dynamic Arctic voyage planning.

\subsection{Computational workflow overview}

\begin{figure}[t!]
 \centering
 \includegraphics[width=0.9\linewidth]{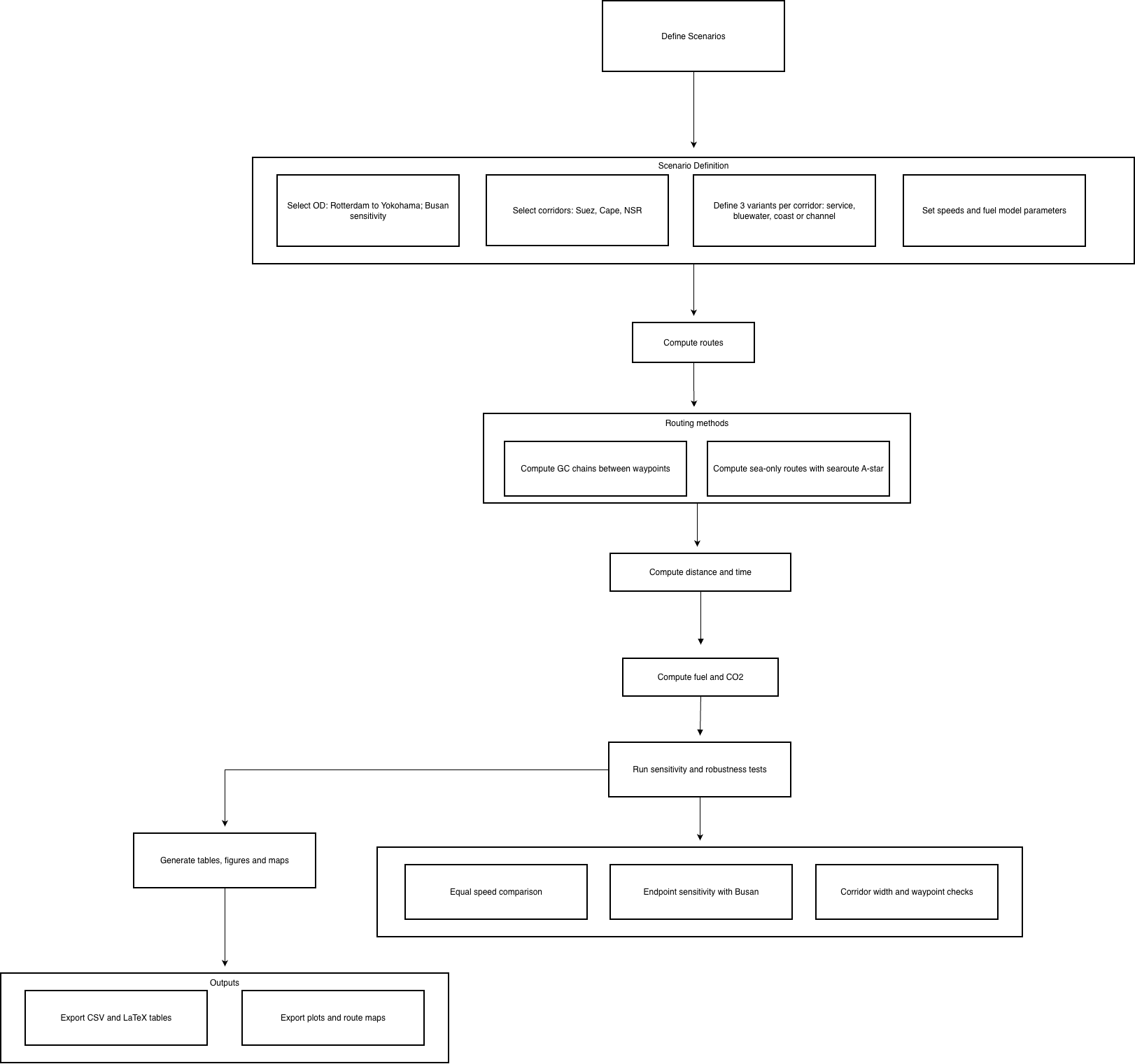}
 \caption{Overview of the methodological pipeline used to compute GC and sea-only routes, 
 derive time/fuel/CO$_2$ metrics, and generate reproducible tables and maps.}
 \label{fig:pipeline}
\end{figure}

Figure~\ref{fig:pipeline} summarises the computational pipeline. 
Predefined macro-waypoints feed into GC and sea-only (A*) routing, from which 
distance--time--fuel--CO$_2$ metrics are derived under corridor-specific speeds. 
All intermediate outputs (routes, metrics, tables, and maps) are exported for 
reproducibility and further analysis.

\FloatBarrier

\section{Results and discussion}\label{sec:results}

\subsection{Case 1: Distance baselines and routing-method effects}\label{sec:case01_dist}

\begin{figure}[t]
  \centering
  \includegraphics[width=.72\linewidth]{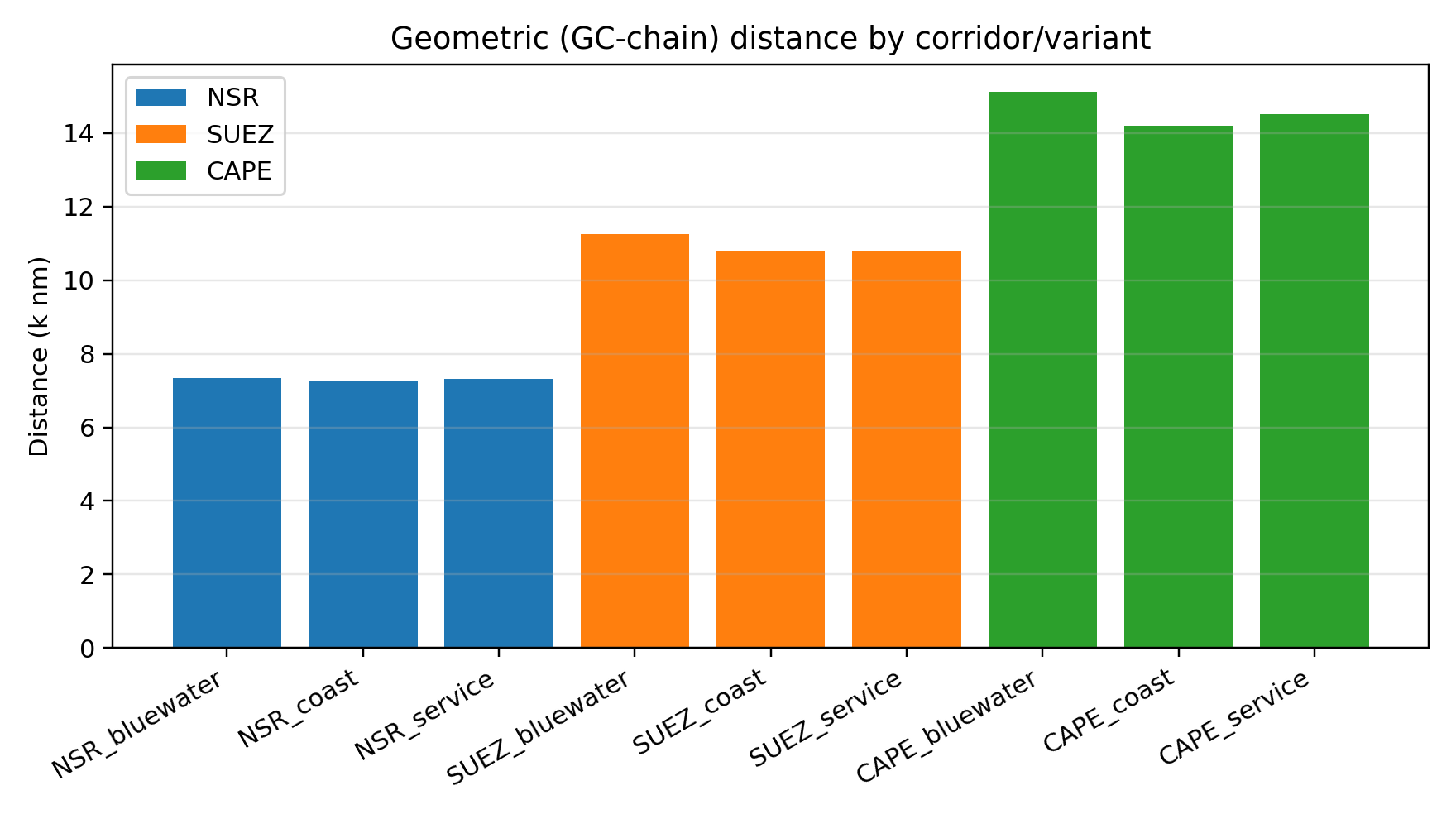}
  \caption{Geometric GC-chain corridor distances by corridor and waypoint
  variant for Rotterdam--Yokohama.\label{fig:gc_bars}}
\end{figure}

\begin{figure}[t]
  \centering
  \includegraphics[width=\linewidth]{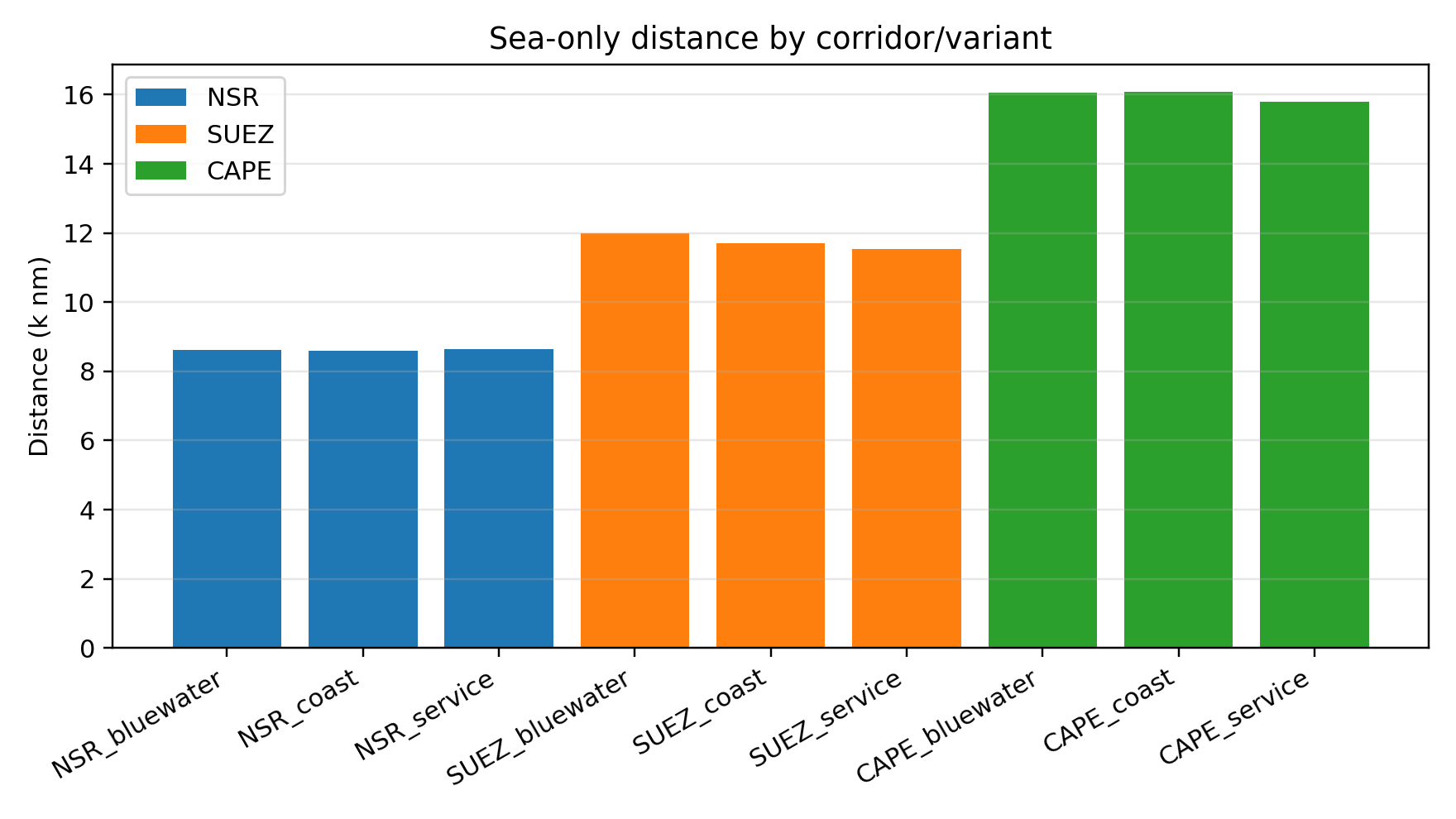}
  \caption{Sea-only (A*) distance by corridor and waypoint variant. The
  Northern Sea Route remains shortest, followed by Suez and Cape.\label{fig:sea_dist}}
\end{figure}

Enforcing sea-only feasibility preserves the qualitative distance ranking 
Northern Sea Route < Suez < Cape of Good Hope across all waypoint 
variants, as shown in Figures ~\ref{fig:gc_bars} and ~\ref{fig:sea_dist}. Per-variant metrics are reported 
in Table~\ref{tab:baseline_all}, and corridor-level medians in Table~\ref{tab:baseline_medians}.

For Suez and Cape, GC-chain and sea-only distances remain close,
reflecting mostly open-ocean legs. For the Northern Sea Route, the sea-only
constraint introduces a larger adjustment because routing must respect Arctic
coastlines, straits, and the Bering gateway.

\begin{figure}[t]
  \centering
  \includegraphics[width=0.9\linewidth]{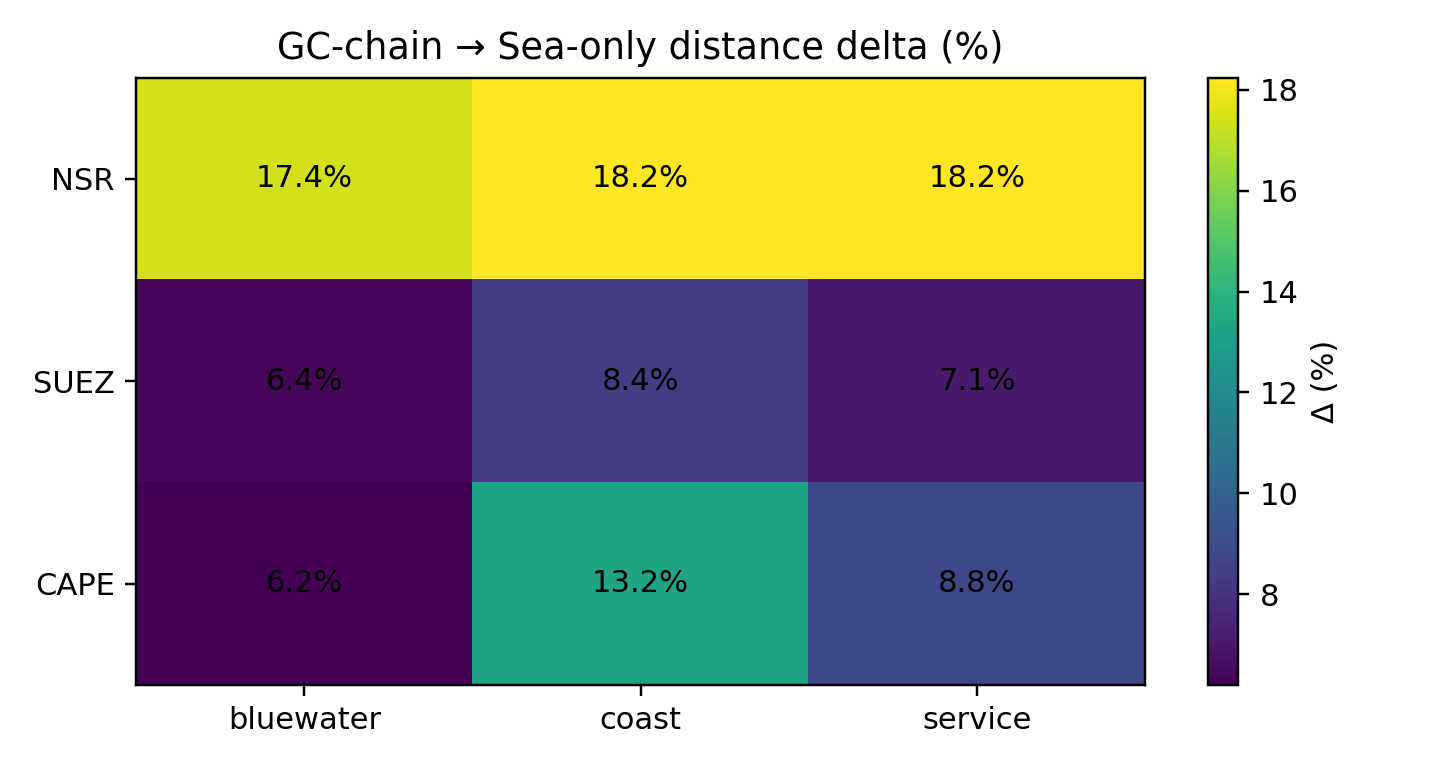}
  \caption{GC-chain to sea-only distance delta (\%) by corridor and waypoint
  philosophy.\label{fig:delta_heatmap}}
\end{figure}

GC-chain to sea-only deltas are modest for Suez
(\(\sim\)6.4--8.4\%) and Cape (\(\sim\)6.2--13.2\%), but substantially larger
for the Northern Sea Route (\(\sim\)17.4--18.2\%) (Figure~\ref{fig:delta_heatmap}).
These deltas confirm that geometry-only baselines materially underestimate
navigable Northern Sea Route length, while having a smaller influence on the
lower-latitude corridors.

Sea-only distances (Figure~\ref{fig:sea_dist}) and their medians Table~\ref{tab:baseline_medians} confirm that the Northern Sea Route remains
shortest by nautical miles. However, for this Rotterdam--Yokohama pair the
headline geometric advantage quoted in the literature (often \(\sim\)30--40\%
from pure geometry) \emph{shrinks materially} once sea-only feasibility and
realistic corridor geometries are enforced, yielding an effective advantage of
roughly \(\sim\)25--30\% versus Suez depending on variant.

Within-corridor spread across the three waypoint philosophies is small
(Figures~\ref{fig:gc_bars} and \ref{fig:sea_dist}); the distance coefficient of
variation is only a few percent (median CV around 2--3\%), indicating that the
main findings are not an artefact of a single waypoint choice.

\begin{figure}[t]
  \centering
  \includegraphics[width=\linewidth]{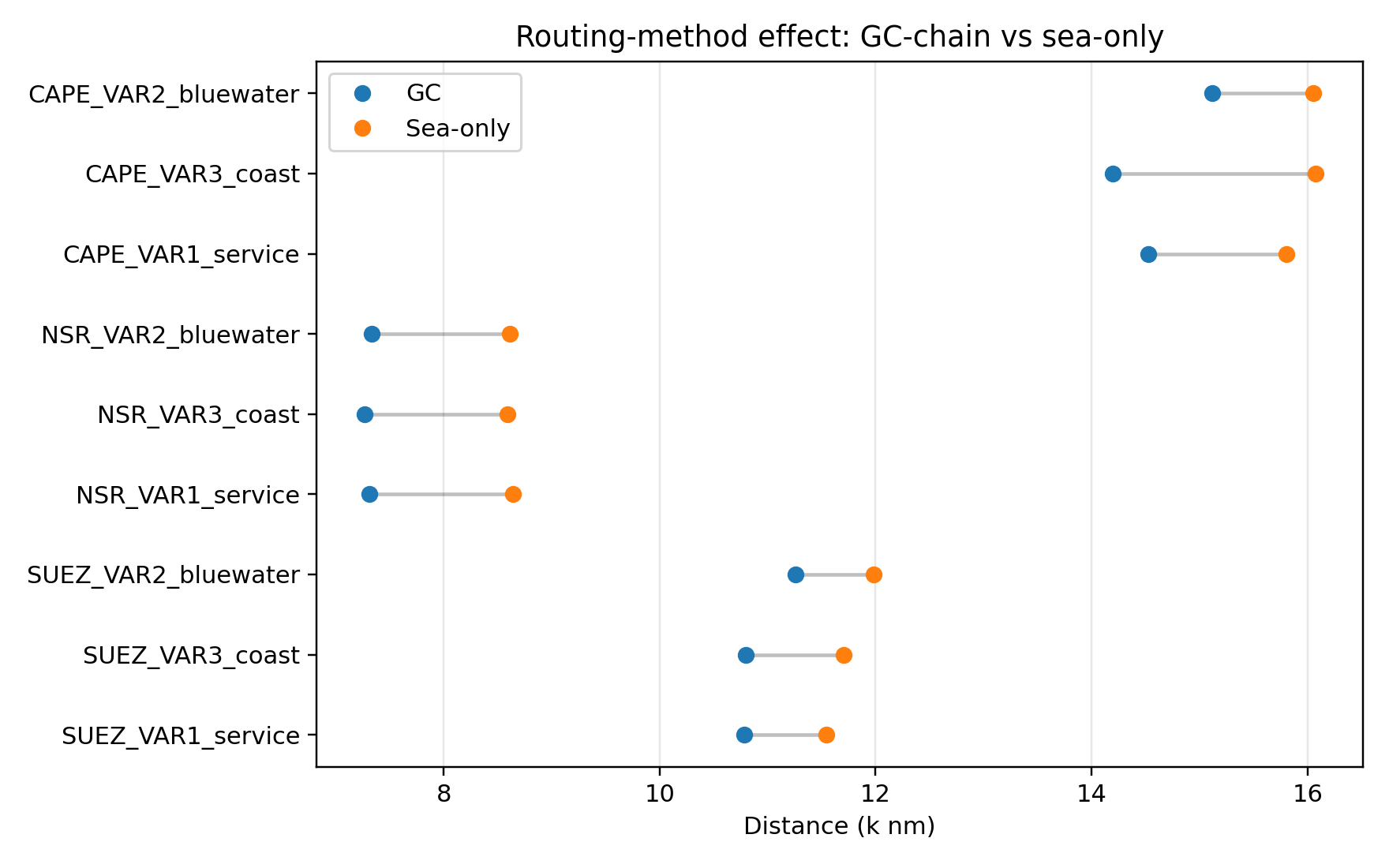}
  \caption{Routing-method effect on distance: GC-chain vs sea-only per
  corridor/variant.\label{fig:gc_vs_sea}}
\end{figure}

\begin{table}[t]
  \centering
  \caption{Baseline metrics per variant: GC-chain vs sea-only distance, and
  derived indicative time and fuel/CO\textsubscript{2}.}
  \label{tab:baseline_all}
    \resizebox{\linewidth}{!}{%
    \begin{tabular}{lllrrrr}
\toprule
corridor & Variant & variant & GC (k\,nm) & Sea-only (k\,nm) & $\Delta$ (nm) & GC$\to$Sea $\Delta$ (\%) \\
\midrule
CAPE & CAPE\_VAR1\_service   & service   & 14.52 & 15.80 & 1278.70 & 8.80 \\
CAPE & CAPE\_VAR2\_bluewater & bluewater & 15.11 & 16.05 &  937.61 & 6.20 \\
CAPE & CAPE\_VAR3\_coast     & coast     & 14.20 & 16.07 & 1876.24 & 13.20 \\
NSR  & NSR\_VAR1\_service    & service   &  7.31 &  8.64 & 1333.23 & 18.20 \\
NSR  & NSR\_VAR2\_bluewater  & bluewater &  7.34 &  8.61 & 1276.45 & 17.40 \\
NSR  & NSR\_VAR3\_coast      & coast     &  7.27 &  8.59 & 1323.14 & 18.20 \\
SUEZ & SUEZ\_VAR1\_service   & service   & 10.78 & 11.55 &  764.01 & 7.10 \\
SUEZ & SUEZ\_VAR2\_bluewater & bluewater & 11.26 & 11.98 &  722.94 & 6.40 \\
SUEZ & SUEZ\_VAR3\_coast     & coast     & 10.80 & 11.70 &  905.80 & 8.40 \\
\bottomrule
\end{tabular}
  }
\end{table}

\begin{table}[t]
  \centering
  \caption{Sea-only distance medians by corridor across all waypoint variants
  (Rotterdam--Yokohama).}
  \label{tab:baseline_medians}
  \begin{tabular}{lrrrrr}
\toprule
corridor &  Median (k nm) &  Min (k nm) &  Max (k nm) &    std &  CV (\%) \\
\midrule
CAPE & 16.05 & 15.80 & 16.07 & 150.27 & 0.90 \\
NSR  &  8.61 &  8.59 &  8.64 &  27.20 & 0.30 \\
SUEZ & 11.70 & 11.55 & 11.98 & 220.65 & 1.90 \\
\bottomrule
\end{tabular}
\end{table}

\subsection{Case 2: Downstream propagation to time, fuel, and CO$_2$ under corridor-typical speeds}
\label{sec:case2}

Case~1 showed that enforcing sea-only feasibility preserves the distance ordering
(NSR $<$ SUEZ $<$ CAPE), while revealing non-trivial GC$\rightarrow$sea-only
corrections. Case~2 addresses the paper’s secondary objective by tracing how
these sea-only distance baselines propagate into indicative voyage time and the
resulting fuel and CO$_2$ estimates under corridor-typical operating speeds.

\begin{figure}[t]
  \centering
  \includegraphics[width=\linewidth]{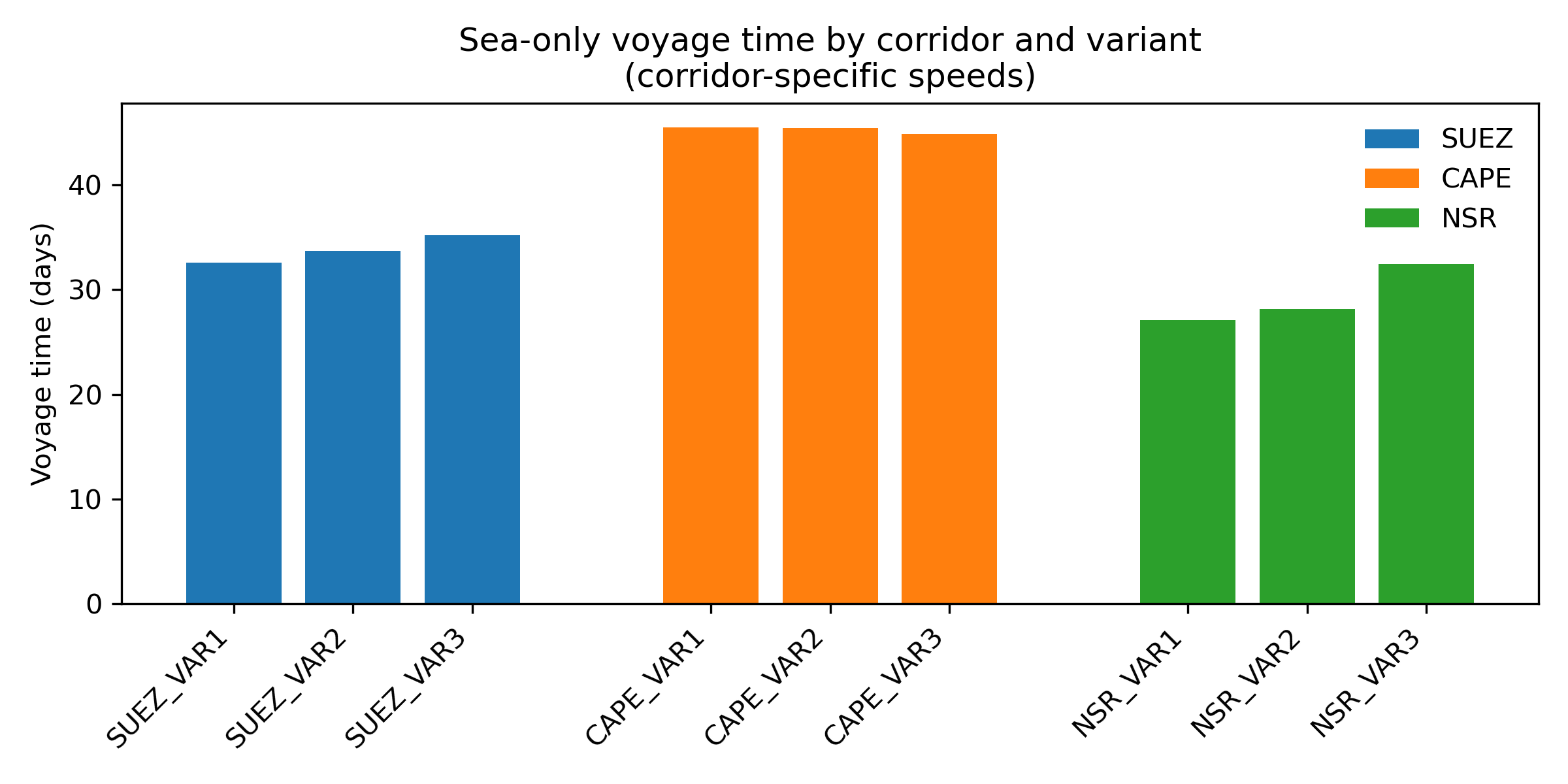}
  \caption{Voyage duration by corridor and variant for sea-only
  routes under corridor-typical service speeds. NSR remains shortest by
  distance, but its time advantage over SUEZ compresses under the lower NSR
  speed.}
  \label{fig:sea_days}
\end{figure}

Sea-only distances are converted to days at sea using representative
corridor speeds (NSR $\approx 12.5$~kn; SUEZ $\approx 14.5$~kn; CAPE
$\approx 14$~kn; Table~\ref{tab:params}). Figure~\ref{fig:sea_days} shows that
distance does not translate linearly into schedule outcomes once realistic
speed policies are applied. Although NSR remains shortest by nautical miles,
its lower corridor-typical speed compresses the schedule advantage relative to
SUEZ. Across variants, indicative durations cluster around
$\sim$29~days for NSR, $\sim$33--35~days for SUEZ, and $\sim$47--48~days for CAPE,
with only modest within-corridor spread.
Thus, a nominal 25–30\% NSR distance advantage corresponds to only a 15\% time advantage in this fair-weather baseline.

\begin{figure}[t]
  \centering
  \includegraphics[width=.85\linewidth]{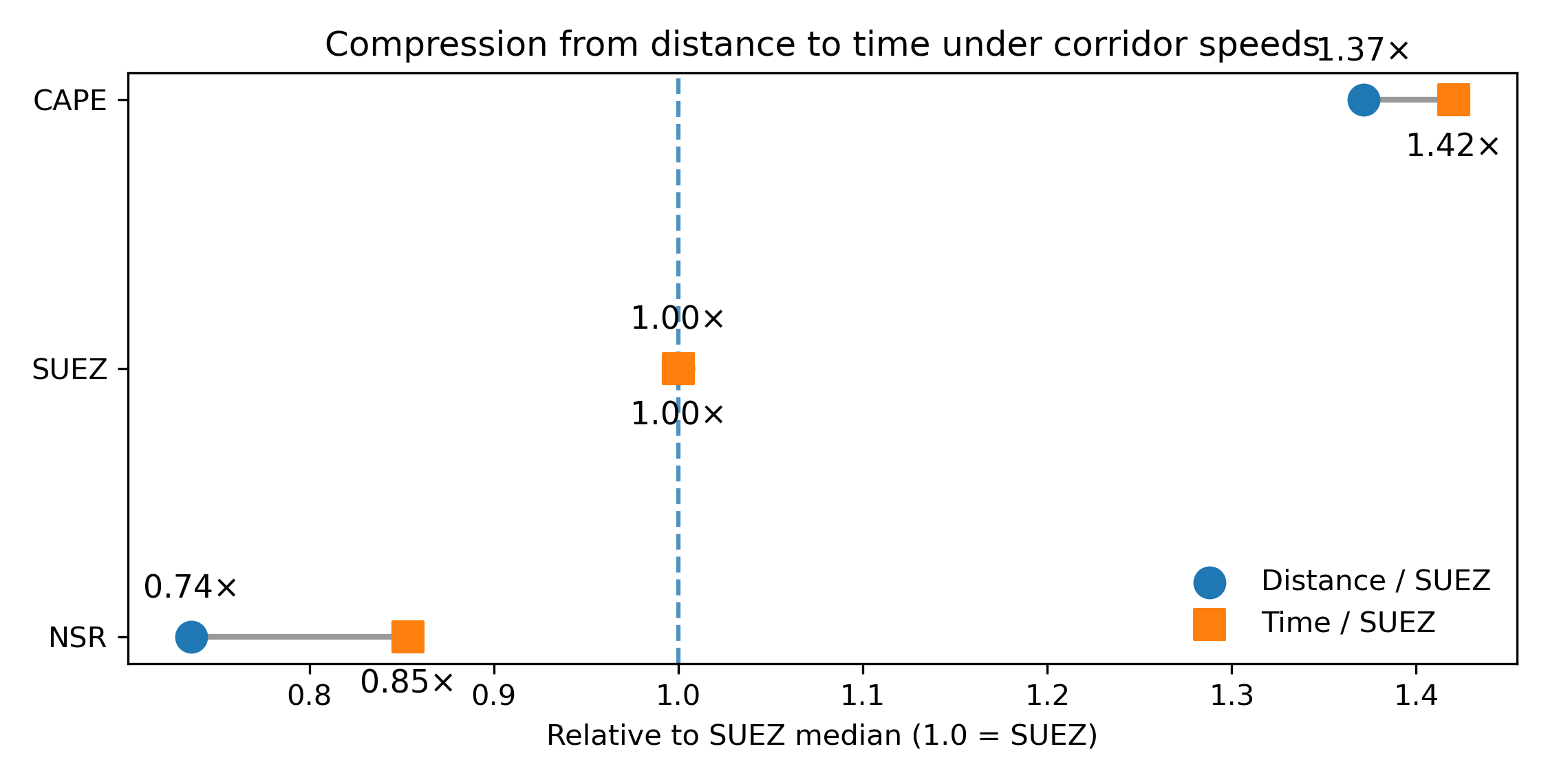}
  \caption{Compression from distance to time under corridor-typical speeds.
  Points show corridor medians across variants, normalised by SUEZ.
  NSR’s distance advantage translates into a smaller time advantage once
  realistic speeds are applied.}
  \label{fig:dist_time_compression}
\end{figure}

The compression is made explicit in Figure~\ref{fig:dist_time_compression},
which compares corridor medians normalised by SUEZ. NSR’s median sea-only
distance is about $0.74\times$ SUEZ, but the corresponding time is only
$0.85\times$ SUEZ. Conversely, CAPE is about $1.37\times$ SUEZ in distance and
$1.42\times$ in time. This directly supports the core message of the paper:
\emph{shorter distance is an optimistic proxy for operational advantage unless
speed policy is stated explicitly}.

\begin{figure}[t]
  \centering
  \includegraphics[width=.9\linewidth]{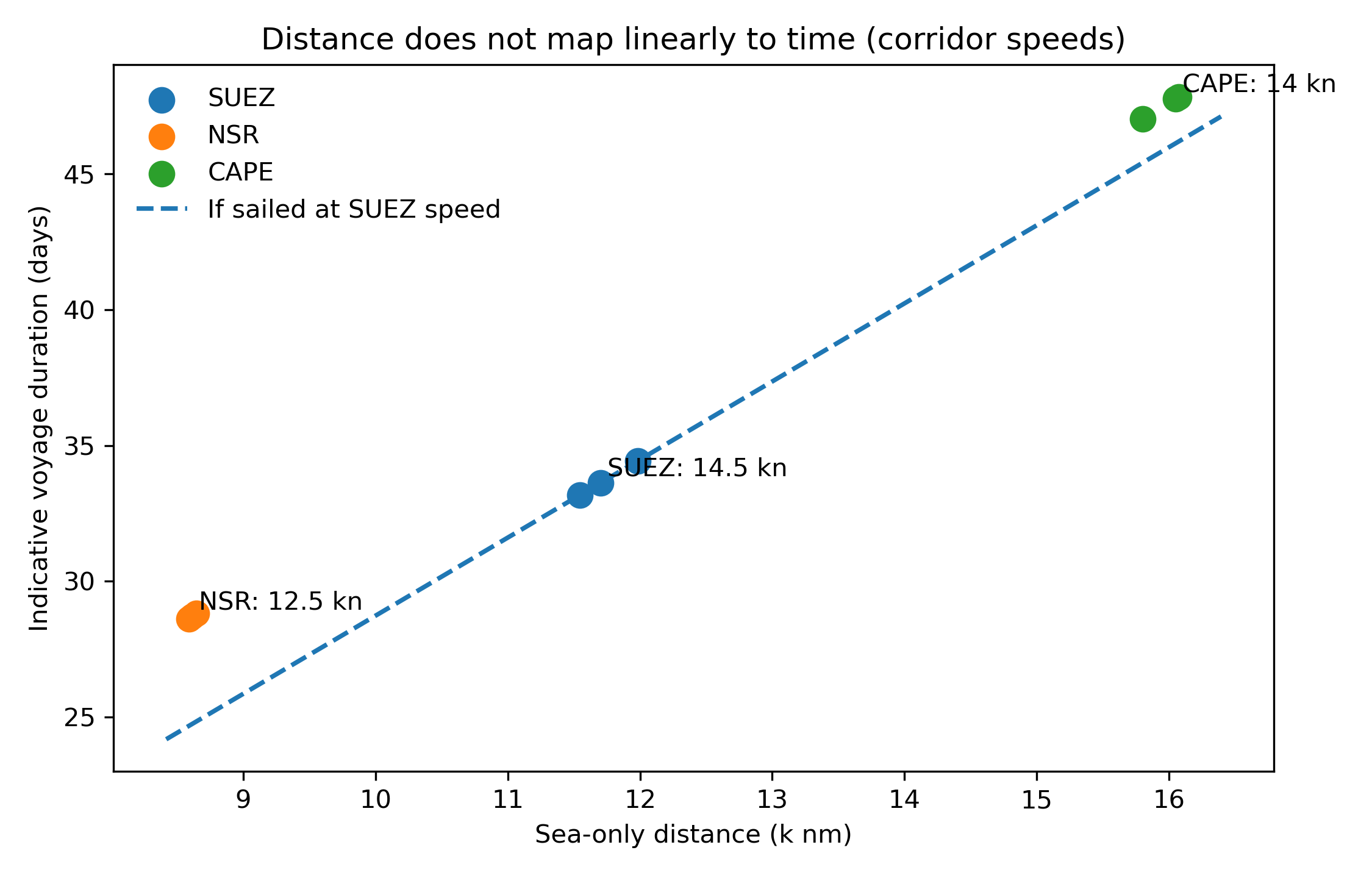}
  \caption{Sea-only distance vs.\ indicative voyage duration under corridor
  speeds. The dashed line shows the duration each route would have at SUEZ
  speed. NSR points lie above this line because NSR speed is lower.}
  \label{fig:dist_vs_days}
\end{figure}

Figure~\ref{fig:dist_vs_days} provides a complementary view: if all corridors
sailed at SUEZ speed (dashed reference line), NSR points would fall on the same
distance–time trend, but under NSR’s lower speed they sit above that line. This
visualises the non-linear distance–time mapping that motivates the remainder of
the analysis.

\begin{figure}[t]
  \centering
  \includegraphics[width=.85\linewidth]{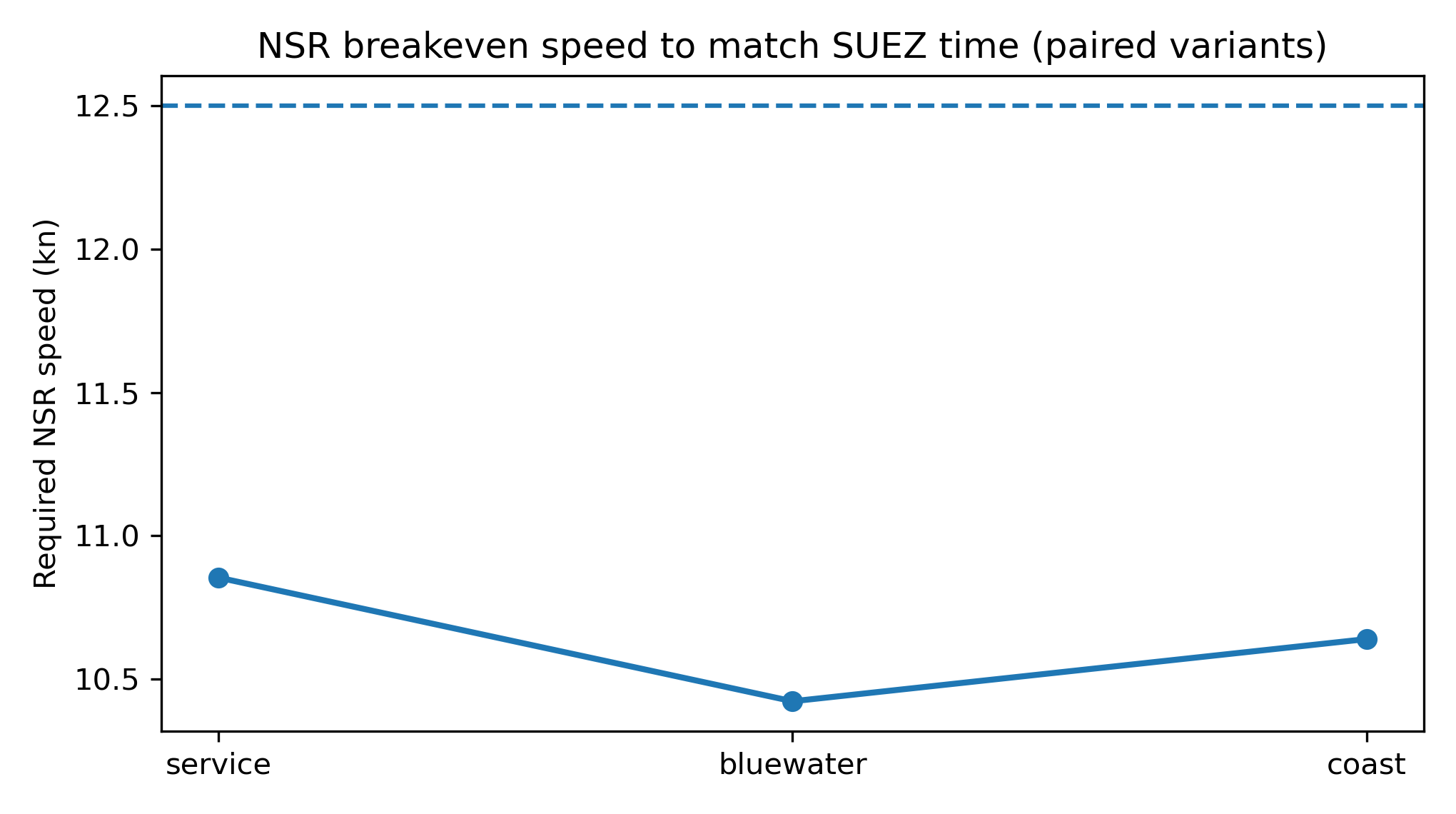}
  \caption{NSR break-even speed required to match SUEZ voyage time for paired
  variants. If NSR average speed drops below this band, its schedule advantage
  disappears.}
  \label{fig:nsr_breakeven}
\end{figure}

To quantify the speed threshold behind this compression, the break-even
analysis in Figure~\ref{fig:nsr_breakeven} shows the NSR service speed required
to match SUEZ duration for each paired variant. The required NSR speed is
$\sim$10.4--10.9~kn (median $\sim$10.7~kn). Thus, for Rotterdam--Yokohama the NSR
remains schedule-advantaged at the assumed 12.5~kn, but \emph{only modestly};
a reduction of NSR average speed below the break-even band would remove the time
advantage entirely.

\begin{figure}[t]
  \centering
  \includegraphics[width=\linewidth]{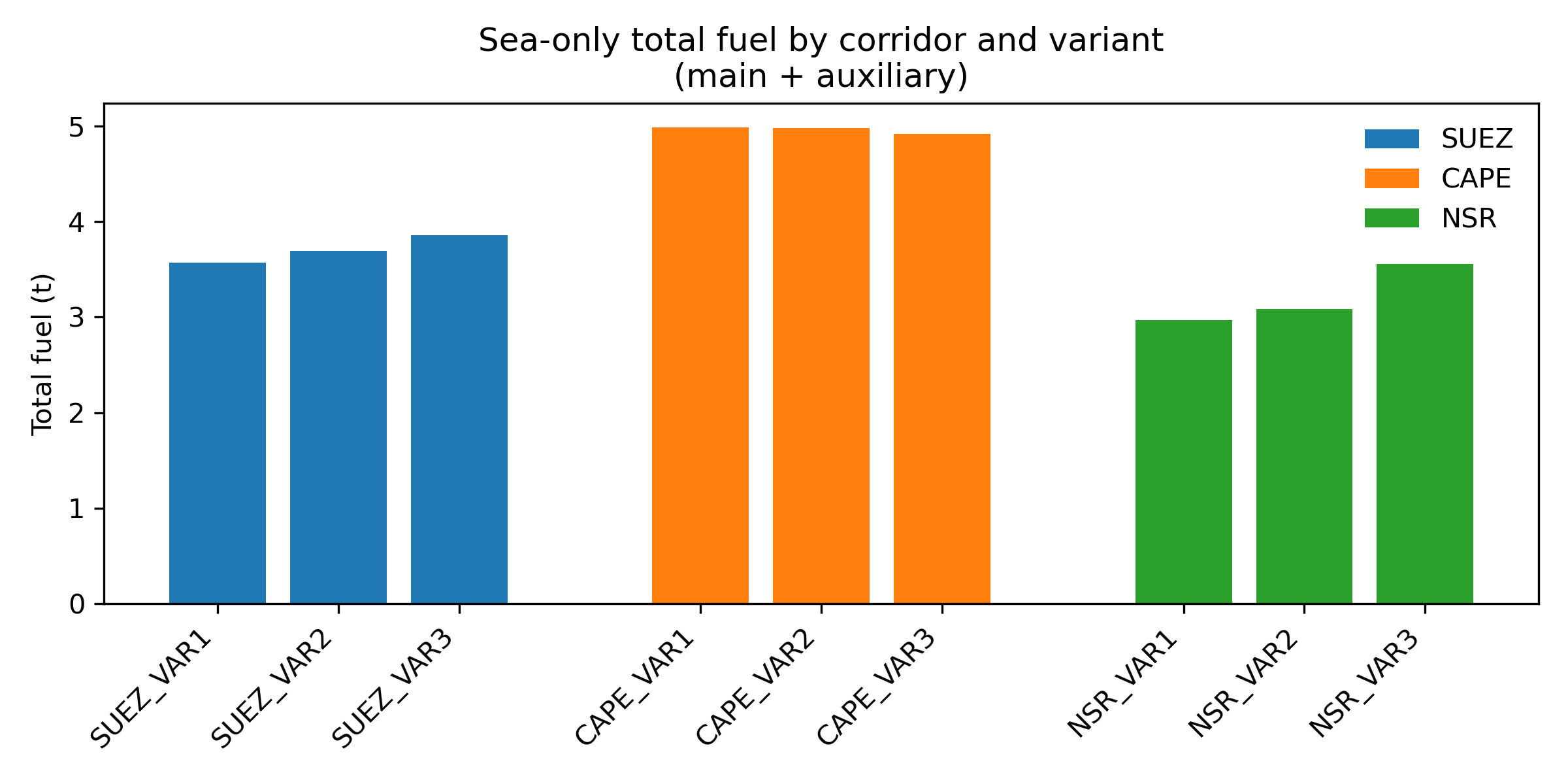}
  \caption{Estimated total fuel consumption by corridor and variant for
  sea-only routes (main + auxiliary). CO$_2$ follows the same pattern because
  it scales with total fuel.}
  \label{fig:sea_fuel}
\end{figure}

Fuel and emissions are computed from voyage hours using the transparent
main- plus auxiliary-engine accounting in Section~\ref{sec:energy_model}.
Figure~\ref{fig:sea_fuel} shows that total fuel (and thus CO$_2$) follows time
rather than distance alone. Absolute totals preserve the ordering
NSR $<$ SUEZ $<$ CAPE, but the NSR margin over SUEZ is smaller than the
distance gap because corridor-typical speeds reduce the hours-at-sea advantage.
Indicative totals are approximately $\sim$0.92~kt for NSR,
$\sim$1.9~kt for SUEZ, and $\sim$2.4~kt for CAPE, implying roughly
$\sim$2.9, 5.9, and 7.5~kt CO$_2$ using the standard factor.
Within-corridor variation across waypoint philosophies remains small, showing
that the downstream conclusions are robust to plausible corridor realisations.

Overall, Case~2 demonstrates the downstream consequence of routing-method
choice: geometric distance advantages provide useful lower bounds, but once
converted to operational metrics under corridor-typical speeds,
the NSR advantage for this Europe--Asia origin–destination pair compresses
substantially. This completes the study objective by showing how routing
representation and operating policy jointly determine corridor competitiveness
in time, fuel, and emissions.

\subsection{Case 3: Sensitivity and robustness}\label{sec:case4}

Case~3 tests whether the distance and downstream conclusions from
Cases~1--2 remain stable under plausible changes in endpoint location and
operating-speed policy.

\begin{figure}[t]
  \centering
  \includegraphics[width=.9\linewidth]{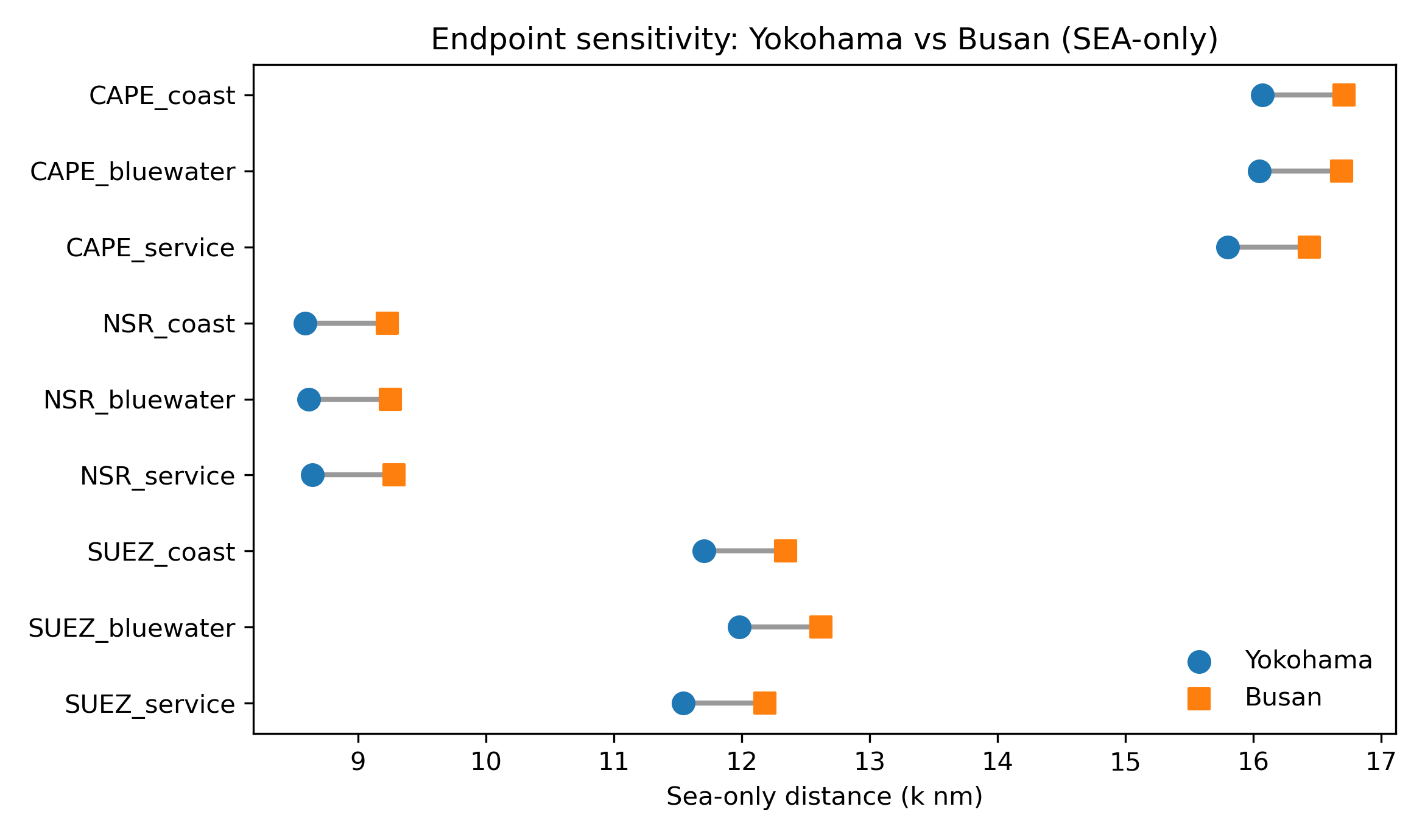}
  \caption{Endpoint sensitivity for sea-only distances: Yokohama versus Busan
  by corridor and waypoint variant. All corridors lengthen by $\approx 0.64$~k~nm,
  with the largest relative increase for NSR.}
  \label{fig:endpoint_sensitivity}
\end{figure}

Figure~\ref{fig:endpoint_sensitivity} compares sea-only distances when the
East Asia endpoint shifts from Yokohama to Busan. The shift lengthens all
corridors by a similar absolute amount of about $0.64$~k~nm, reflecting a
northwestward relocation of the destination. However, the \emph{relative}
impact differs by corridor: the Northern Sea Route increases by
$\approx 7.39$--$7.44\%$ across variants, compared with $\approx 5.33$--$5.53\%$
for Suez and $\approx 3.97$--$4.04\%$ for Cape. This pattern is consistent with
the corridor geometry: moving to Busan shortens the northwest Pacific approach
from the Bering Strait, so the NSR retains (and slightly enlarges) its distance
advantage over Suez for more northerly/westerly East Asian markets.
Importantly, the qualitative ordering is unchanged (NSR $<$ SUEZ $<$ CAPE),
showing that the main distance ranking is not an artefact of a single
destination choice.

\begin{figure}[t]
  \centering
  \includegraphics[width=.9\linewidth]{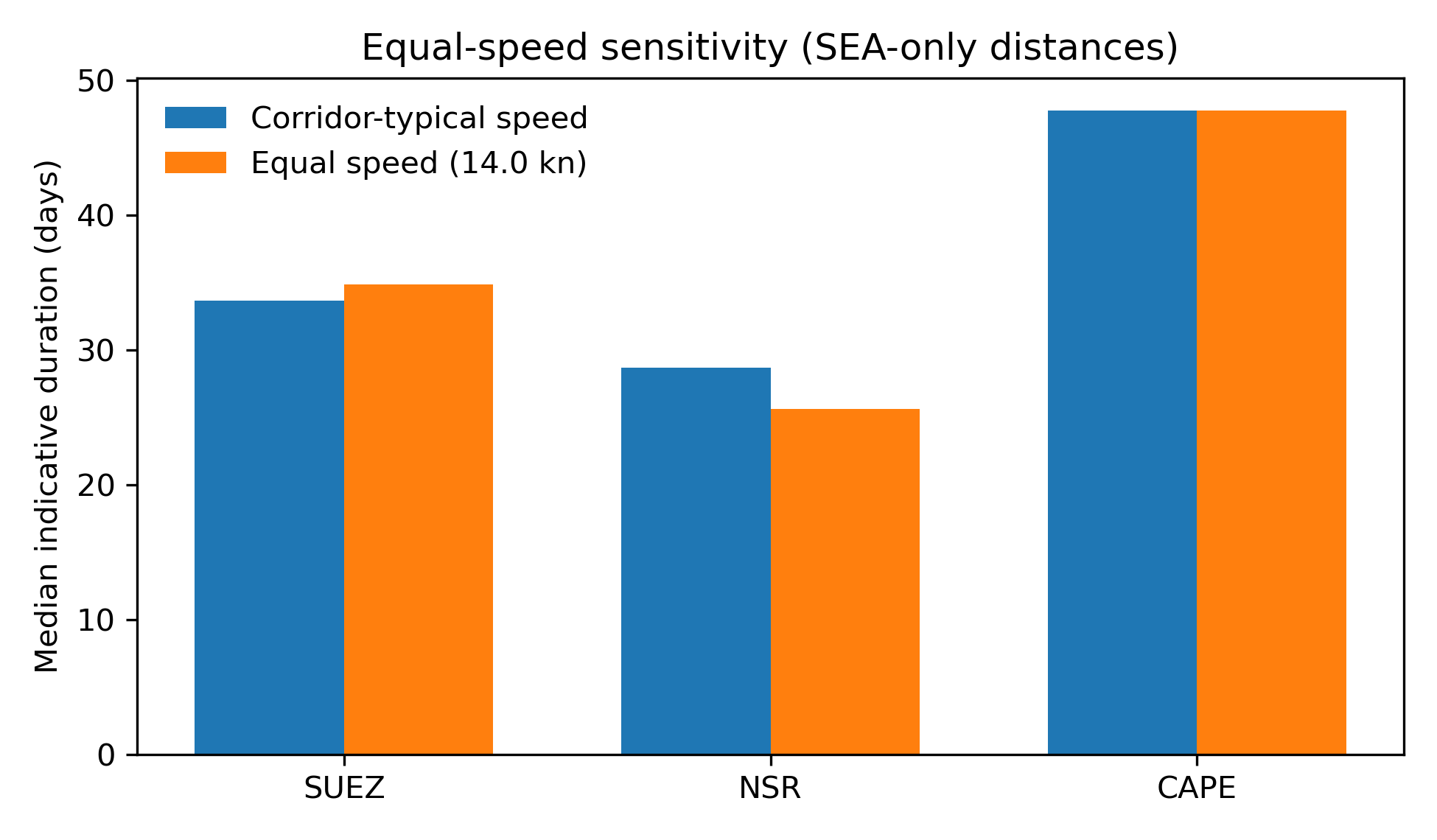}
  \caption{Equal-speed sensitivity for sea-only routes. Median durations under
  corridor-typical speeds are compared to an equal-speed scenario, isolating
  operating-policy effects from geometry.}
  \label{fig:equal_speed_days}
\end{figure}

To isolate pure geometry from operating policy, we recomputed indicative voyage
times using a common service speed across corridors. As shown in
Figure~\ref{fig:equal_speed_days}, equal-speed assumptions restore an almost
linear distance-to-time mapping and recover the geometric ordering from Case~1.
The comparison makes explicit that the compression of NSR’s schedule and fuel
advantages in Case~2 is driven by corridor-typical speed differences rather
than by instability in the routing baselines.

Overall, Case~3 confirms that the corridor rankings and the key downstream
interpretations are robust to reasonable endpoint relocation and to the choice
between corridor-typical and equal-speed operating assumptions, directly
supporting the paper’s contribution of isolating routing-method effects
from operational and geographic confounders.

Additional tests with perturbed corridor widths and macro-waypoint locations (not shown) changed absolute sea-only distances by at most a few percent and did not affect the NSR < SUEZ < CAPE ranking or the main time–fuel–CO$_2$ patterns.

\subsection{Implications and limitations}\label{sec:implications}

For corridor selection and fleet planning, three points stand out:
(i) \emph{Corridor choice and speed policy} matter as much as raw distance; 
distance-based comparisons alone can materially overstate the NSR’s practical 
schedule and fuel advantage over SUEZ for the Rotterdam--Yokohama trade. 
(ii) Equal-speed comparisons are useful for isolating geometry, but 
\emph{corridor-typical speeds} determine realised schedules, fuel use, and 
auxiliary exposure. 
(iii) The CAPE corridor, although longest, remains a robust year-round fallback 
that is independent of canals and Arctic seasonality. These implications are 
stable under plausible endpoint relocation (Busan vs.\ Yokohama) and under 
equal-speed sensitivity tests (Case~3).

By construction, the present analysis \emph{excludes} dynamic environmental and 
regulatory constraints (sea ice, winds, waves, currents, emission-control areas, 
canal queues, ice-pilotage and escort requirements). These are essential for 
operational decision-making but are deliberately omitted here to isolate 
routing-method effects. The fuel and CO$_2$ model is transparent but not 
ship-specific: it does not include detailed resistance models, 
speed--power--SFOC coupling, or weather-dependent added resistance. Finally, the 
results are computed for a single origin--destination pair with curated 
waypoints; the framework is general, but quantitative deltas will vary for 
other trades and waypoint sets.

Overall, the results show that routing representation is a first-order driver 
of corridor conclusions. Geometry-only reasoning reproduces the familiar 
``Arctic shortcut'' narrative, whereas conservative sea-only routing combined 
with corridor-typical speeds reveals a much more compressed and corridor- 
dependent picture of time, fuel, and emissions.

These fair-weather baselines provide the routing layer for subsequent work in which sea ice, metocean forcing, and regulatory overlays are introduced on top of the sea-only A* framework.
\section{Conclusions}\label{sec:conclusions}

This paper quantified how routing representation---geometric great-circle (GC) 
versus physically feasible sea-only paths---affects distance, time, fuel, and 
CO$_2$ estimates for Europe--Asia trade across three strategic corridors: 
the Suez Canal, the Cape of Good Hope, and the Arctic Northern Sea Route (NSR). 
A coastline-masked A* framework was applied to Rotterdam--Yokohama with three 
operational waypoint philosophies per corridor and a transparent main- and 
auxiliary-engine accounting.

Three main conclusions emerge:
\begin{enumerate}
  \item \textbf{Sea-only routing preserves qualitative distance ordering but alters magnitude.} 
  The ranking NSR $<$ SUEZ $<$ CAPE holds once navigability is enforced, with 
  small GC$\rightarrow$sea-only adjustments for SUEZ and CAPE and larger adjustments 
  for NSR driven by Arctic coastal geometry and chokepoints. As a result, NSR’s 
  widely cited geometric advantage shrinks for this OD pair under feasible routing.

  \item \textbf{Distance does not translate linearly into time, fuel, or CO$_2$.} 
  Under corridor-typical speeds, NSR’s schedule and fuel advantages compress 
  and may disappear relative to SUEZ. CAPE remains longest in both distance and 
  duration. Fuel and CO$_2$ primarily follow hours at sea, underscoring that 
  distance alone is an insufficient proxy for operational or environmental performance.

  \item \textbf{Competitiveness depends on endpoint and speed policy.} 
  NSR’s relative advantage increases for more northerly East Asian ports 
  (e.g., Busan) and decreases for Yokohama. Equal-speed comparisons recover 
  pure geometric differences, while corridor-typical speeds show how operating 
  practices can offset or amplify them. CAPE emerges as a robust canal-independent fallback rather than a competitive primary option.
\end{enumerate}

Methodologically, the study provides a reproducible, corridor-agnostic baseline 
linking routing-method choice to distance, time, fuel, and CO$_2$ for 
Europe--Asia services. The findings are robust across waypoint philosophies and 
sensitivity tests, and the framework is designed for extension. Future work 
will integrate seasonal and daily sea-ice fields, metocean routing, 
emission-control areas and canal constraints, and ship-specific propulsion models, 
extending the static fair-weather baseline toward dynamic voyage-planning and 
digital-twin applications for Arctic and global shipping.
\section*{CRediT authorship contribution statement}
\textbf{Abdella Mohamed:} Conceptualization, Methodology, Coding, Data curation, Writing, Visualization, Investigation, Validation. 
\textbf{Xiangyu Hu:} Supervision, Conceptualization, Methodology, Manuscript review and editing. 
\textbf{Christian Hendricks:} Mentorship, Conceptualization, Methodology, Manuscript review and editing. 
\section*{Acknowledgements}
This work was supported by Technical University of Munich and Everllence (formerly: MAN Energy Solutions).

\section*{Data and code availability}

All code and data required to reproduce the routing, distance, time, fuel, and
CO$_2$ results in this paper are openly available via Zenodo under the DOI
\href{https://doi.org/10.5281/zenodo.17767072}{10.5281/zenodo.17767072}.
The repository snapshot archived there contains:

\begin{itemize}
  \item the Python package used in this study (routing pipeline, corridor and
        waypoint definitions, A* / sea-only distance computation, and
        post-processing utilities);
  \item Jupyter notebooks organised by case (distance baselines, time/fuel/CO$_2$
        propagation, endpoint and equal-speed sensitivity) that reproduce all
        tables and figures in the Results and discussion section;
  \item static input files (macro-waypoint manifests, corridor masks, and
        pre-computed route geometries) and parameter tables corresponding to
        the assumptions documented in Section~\ref{sec:data_methods};
  \item export-ready figure files used in the manuscript.
\end{itemize}

The Zenodo record is linked to the underlying GitHub repository, where ongoing
development and issue tracking take place. Users are encouraged to cite the
Zenodo DOI when reusing the code or data. External dependencies such as the
\texttt{searoute} library \citep{searoute2022} are referenced separately in the
bibliography.

\section*{Declaration of generative AI and AI-assisted technologies in the manuscript preparation process}

The authors used OpenAI ChatGPT to assist with language editing during 
manuscript preparation. The authors reviewed and edited the content 
and accept full responsibility for the final manuscript.

\listoftables
\listoffigures
\clearpage
\bibliography{main}
\newpage

\setcounter{table}{0}
\end{document}